\documentclass{aa}
\usepackage[varg]{txfonts}
\usepackage{natbib}
\usepackage{graphicx}
\usepackage{subcaption}
\usepackage{wasysym} 
\usepackage[usenames,dvips]{color} 
\usepackage{verbatim} 
\usepackage{rotating}
\usepackage{multirow} 
\usepackage{soul} 
\usepackage{xcolor} 
\usepackage{framed} 
\colorlet{shadecolor}{blue!20} 
\usepackage{centernot} 
\usepackage{amsmath}
\def\phs{\phantom{$-$}}   
\def\phn{\phantom{0}}

\usepackage{array}

\def\apjl{ApJ}%
\def\apjs{ApJS}%
\def\ao{Appl.~Opt.}%
\def\aap{A\&A}%
\title{Applying the expanding photosphere and standardized candle methods to Type II-Plateau supernovae at cosmologically significant redshifts:}
\subtitle{the distance to SN~2013eq\thanks{Spectra of SN 2013eq are available in the Weizmann Interactive Supernova data REPository (WISeREP): http://wiserep.weizmann.ac.il.}}

\author{E.E.E. Gall \inst{\ref{inst1},\ref{inst2}\thanks{E-mail: egall01@qub.ac.uk}}
\and R. Kotak \inst{\ref{inst1}}  
\and B. Leibundgut \inst{\ref{inst3},\ref{inst4}}  
\and S. Taubenberger \inst{\ref{inst2},\ref{inst3}}
\and W. Hillebrandt \inst{\ref{inst2}}  
\and M. Kromer \inst{\ref{inst5}}  
}

\institute{Astrophysics Research Centre, School of Mathematics and Physics, Queen's University Belfast, Belfast BT7 1NN, UK\label{inst1}
\and Max-Planck-Institut für Astrophysik, Karl-Schwarzschild-Str. 1, DE-85748 Garching-bei-M\"{u}nchen, Germany\label{inst2}
\and ESO, Karl-Schwarzschild-Strasse 2, DE-85748 Garching-bei-M\"{u}nchen, Germany\label{inst3}
\and Excellence Cluster Universe, Technische Universität München, Boltzmannstrasse 2, DE-85748 Garching-bei-M\"{u}nchen, Germany\label{inst4}
\and The Oskar Klein Centre \& Department of Astronomy, Stockholm University, AlbaNova, SE-106 91 Stockholm, Sweden\label{inst5}
}

\date{Received 18 February 2016 /
Accepted 22 June 2016 }

\abstract{
Based on optical imaging and spectroscopy of the Type II-Plateau SN~2013eq, we present a comparative study of commonly used distance determination methods based on Type II supernovae. The occurrence of SN~2013eq in the Hubble flow ($z$ = 0.041\,$\pm$\,0.001) prompted us to investigate the implications of the difference between ``angular'' and ``luminosity'' distances within the framework of the expanding photosphere method (EPM) that relies upon a relation between flux and angular size to yield a distance. Following a re-derivation of the basic equations of the EPM for SNe at non-negligible redshifts, we conclude that the EPM results in an angular distance. The observed flux should be converted into the SN rest frame and the angular size, $\theta$, has to be corrected by a factor of $(1+z)^2$. Alternatively, the EPM angular distance can be converted to a luminosity distance by implementing a modification of the angular size. 
For SN~2013eq, we find EPM luminosity distances of $D_L$ = 151\,$\pm$\,18\,Mpc and $D_L$ = 164\,$\pm$\,20\,Mpc by making use of  different sets of dilution factors taken from the literature. Application of the standardized candle method for Type II-P SNe results in an independent luminosity distance estimate ($D_L$ = 168\,$\pm$\,16\,Mpc) that is consistent with the EPM estimate. 
}


\begin{document}

\titlerunning{The distance to SN 2013eq}
\maketitle

\section{Introduction}

Supernovae have proven to be useful as distance indicators and are pivotal to estimating fundamental cosmological parameters such as the expansion rate, geometry, age, and energy content of the Universe. Observations using thermonuclear (Type Ia) supernovae (SNe) led to the surprising conclusion that the Universe was expanding at an accelerating rate \citep{Riess1998a,Perlmutter1999a,Leibundgut2001,Goobar2011}.

Although the use of SNe Ia is very well established, it is also possible to use core-collapse SNe \citep{Hamuy2002a}.
In particular, two methods for distance determinations -- using Type II-Plateau (P) SNe -- have received the most attention. 
The first is the so-called ``expanding photosphere method'' (EPM) that was developed by \citet{Kirshner1974} based on a proposition by Leonard Searle, who suggested that the Baade-Wesselink method \citep{Baade1926,Wesselink1946}, used to determine the radii of pulsating stars, could be adapted to estimate distances to SNe. This could be done by linking the photospheric expansion to the observed expansion velocities. Thus the EPM is essentially a geometrical technique relying on the comparison between the angular size of an object and its observed flux. 
Over the past 40 years a variety of improvements have been suggested e.g. the introduction of distance correction factors to adjust for dilution effects in scattering atmospheres, detailed modelling of SN atmospheres, and cross-correlation techniques to measure line velocities \citep[e.g.][]{Wagoner1981,Eastman1996,Hamuy2001,Dessart2005}. A somewhat different, though related form, is the spectral-fitting expanding atmosphere method \citep[SEAM;][]{Baron2004}, in which model fits to the SN spectra are used to determine key variables.

The other commonly used method for distance determination using SNe II-P is the ``standardized candle method'' \citep[SCM;][]{Hamuy2002a}. It rests on the expectation that a more energetic, and consequently more luminous explosion will produce ejecta having a higher kinetic energy per unit mass. This results in a correlation between the bolometric luminosity and the expansion velocities during the plateau phase, allowing for a normalization of the SN luminosity, yielding a distance estimate. A number of groups have further built upon the SCM, for example by simplifying extinction corrections, or exploring alternatives to the commonly employed iron lines to measure the ejecta velocities \citep[e.g.][]{Nugent2006,Poznanski2009}. 
Techniques relying solely on photometric data are also being explored \citep{deJaeger2015}. The obvious advantage of being significantly less demanding in terms of the data required, comes at the cost of lowered accuracy compared to the SCM. Indeed, the above study reports a dispersion of 0.43\,mag for a colour-based distance estimation method, compared to 0.29\,mag for the SCM.

Even though Type II-P SNe are intrinsically fainter than SNe Ia, and thus more challenging to observe at large distances, they occur more frequently per unit volume \citep[e.g.][]{Li2011a}, allowing the possibility of building statistically significant samples.
Moreover, the EPM has the striking virtue that it is independent from local calibrations -- albeit at the cost of requiring multi-epoch spectroscopy alongside photometric observations. The SCM, in comparison, is less observationally expensive requiring mainly data around the midpoint of the plateau phase, but akin to the SN Ia distance determinations, it does rely on local distance anchors. Nevertheless, both methods offer alternative distance estimates, and more importantly, are affected by different systematic effects compared to the SNe Ia. 

In order to create a EPM/SCM Hubble diagram based on Type II-P SNe, distance measurements at and beyond the Hubble flow are essential. Galaxies in the local neighbourhood are affected by peculiar motions that can be difficult to model and therefore limit the precision with which cosmological redshifts can be measured. 

Barring a few exceptions, applications of the EPM or its variations have remained confined to SNe within the local Universe \citep[e.g.,][]{Hamuy2001,Leonard2002b,Elmhamdi2003,Dhungana2015}. 
To our best knowledge the EPM has only been adopted for SNe with redshifts $z > 0.01$ by \citet{Schmidt1994a} who performed the EPM on SN 1992am at $z \sim 0.049$, \citet{Eastman1996} who also included SN 1992am in their sample and \citet{Jones2009} whose sample encompassed SNe with redshifts up to $z = 0.028$.
\citet{Schmidt1994a} were the first to investigate the implications of applying the EPM at higher redshifts.

On the other hand, probably due to the relative ease of obtaining the minimum requisite data, the SCM is much more commonly applied to SNe at all redshifts $0.01 < z < 0.1$ \citep[e.g.][]{Hamuy2002a,Maguire2010a,Polshaw2015} and even to SNe IIP at redshifts $z > 0.1$ \citep{Nugent2006,Poznanski2009,DAndrea2010}.

Motivated by the discovery of SN 2013eq at a redshift of $z$ = 0.041\,$\pm$\,0.001 we undertook an analysis of the relativistic effects that occur when applying the EPM to SNe at non-negligible redshifts. 
As a result, we expand on earlier work by \citet{Schmidt1994a}, who first investigated the implications of high redshift EPM. We wish to ensure that the difference between angular distance and luminosity distance -- that becomes significant when moving to higher redshifts -- is well understood within the framework of the EPM. 

This paper is structured as follows: observations of SN 2013eq are presented in \S\ref{section:Observations_and_data_reduction}; 
we summarize the EPM and SCM methods in \S\ref{section:Methods}; our results are discussed in \S\ref{section:Results_and_discussion}.

\section{Observations and data reduction}
\label{section:Observations_and_data_reduction}

SN 2013eq was discovered on 2013 July 30 \citep{2013CBET3616} and spectroscopically classified as a Type II SN using spectra obtained on 2013 July 31 and August 1 \citep{2013CBET3616}. These exhibit a blue continuum with characteristic P-Cygni line profiles of H$\alpha$ and H$\beta$, indicating that SN 2013eq was discovered very young, even though the closest pre-discovery non-detection was on 2013 June 19, more than 1 month before its discovery \citep{2013CBET3616}. \citet{2013CBET3616} adopt a redshift of 0.042 for SN 2013eq from the host galaxy. We obtained 5 spectra ranging from 7 to 65 days after discovery (rest frame) and photometry up to 76 days after discovery (rest frame).

\subsection{Data reduction}
\label{section:Data_reduction}

Optical photometry was obtained with the Optical Wide Field Camera, IO:O, mounted on the 2m Liverpool Telescope (LT; Bessell-$B$ and -$V$ filters as well as SDSS-$r'$ and -$i'$ filters).
All data were reduced in the standard fashion using the LT pipelines, including trimming, bias subtraction, and flat-fielding.

Point-spread function (PSF) fitting photometry of SN 2013eq was carried out on all images using the custom built {\sc SNOoPY}\footnote{SuperNOva PhotometrY, a package for SN photometry implemented in IRAF by E. Cappellaro; http://sngroup.oapd.inaf.it/snoopy.html.} 
package within {\sc iraf}\footnote{{\sc iraf} (Image Reduction and Analysis Facility) is distributed by the National Optical Astronomy Observatories, which are operated by the Association of Universities for Research in Astronomy, Inc., under cooperative agreement with the National Science Foundation.}. Photometric zero points and colour terms were derived using observations of Landolt standard star fields \citep{Landolt1992a} in the 3 photometric nights and their averaged values where then used to calibrate the magnitudes of a set of local sequence stars as shown in Table \ref{table:sequence_stars} in the appendix and Figure \ref{figure:SN2013eq_sequence_stars} that were in turn used to calibrate the photometry of the SN in the remainder of nights. We estimated the uncertainties of the PSF-fitting via artificial star experiments. An artificial star of the same magnitude as the SN was placed close to the position of the SN. The magnitude was measured, and the process was repeated for several positions around the SN. The standard deviation of the magnitudes of the artificial star were combined in quadrature with the uncertainty of the PSF-fit and the uncertainty of the photometric zeropoint to give the final uncertainty of the magnitude of the SN. 

\begin{figure}
  \centering
    \includegraphics[width=\columnwidth]{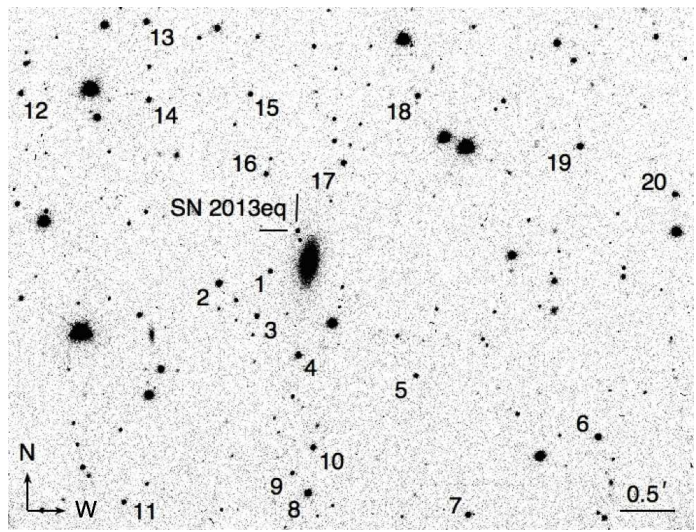}
    \caption{SN 2013eq and its environment. Short dashes mark the location of the supernova at $\alpha_\mathrm{J2000} = 17^{h}33^{m}15^{s}.73$, 
    $\delta_\mathrm{J2000} = +36^{\circ}28'35''.2$. 
     The numbers mark the positions of the sequence stars (see also Table \ref{table:sequence_stars}) used for the photometric calibrations. SDSS-$i'$-band image taken on 2013 August 08, 8.7\,d after discovery (rest frame).}
    \label{figure:SN2013eq_sequence_stars}
\end{figure}

A series of five optical spectra were obtained with the Optical System for Imaging and low-Intermediate-Resolution Integrated Spectroscopy (OSIRIS, grating ID R300B) mounted on the Gran Telescopio CANARIAS (GTC) and the Intermediate dispersion Spectrograph and Imaging System (ISIS, grating IDs R158R and R300B) mounted on the William Herschel Telescope (WHT).

The spectra were reduced using {\sc iraf} following standard procedures. These included trimming, bias subtraction, flat-fielding, optimal extraction, wavelength calibration via arc lamps, flux calibration via spectrophotometric standard stars, and re-calibration of the spectral fluxes to match the photometry.  The spectra were also corrected for telluric absorption using a model spectrum of the telluric bands. 

A weak Na\,{\sc i}\,D absorption, with an equivalent width of EW$_{\mathrm{Na\,I\,D}}$ = 0.547\,$\pm$\,0.072\,\AA, can be detected in the +25\,d spectrum of SN 2013eq. Applying the empirical relation between Na\,{\sc i}\,D absorption and dust extinction given in \citet[][Equation 9]{Poznanski2012}, this translates into an extinction within the host galaxy of $E(B-V)_{\mathrm{host}}$ = 0.062\,$\pm$\,0.028\,mag. 
Even though the EW Na\,{\sc i}\,D absorption is a commonly used diagnostic of extinction, as has been noted on numerous occasions, it is not always reliable \citep[e.g][]{Poznanski2012}.
Given the remote location of SN 2013eq (projected distance of $\sim$\,14.6\,kpc from the host galaxy nucleus (assuming $z$ = 0.041 and $H_0$ = 70\,km\,s$^{-1}$\,Mpc$^{-1}$), the host extinction deduced above is likely to be an upper limit.
Although one expects the local environments of core-collapse SNe to be different from those of SNe Ia, 
there is some indication for a positive correlation between $A_V$ and the radial position of the SN within the host galaxy \citep[e.g.][]{Holwerda2015}. For the Galactic extinction we adopt a value of 
$E(B -V)_\mathrm{{Gal}}$ = 0.034\,mag \citep{Schlafly2011}.

The Na\,{\sc i}\,D absorption lends itself to an estimate of the redshift.
Applying Equation 12 in \citet[][using the values for the rest wavelengths and oscillator strengths of the individual lines as given in their Table 4]{Leonard2002a}, and taking the rest wavelength of the Na\,{\sc i}\,D $\lambda\lambda$5890,5896 multiplet to be $\lambda_\mathrm{Na\,I\,D}$ = 5891.94\,\AA, a comparison with the observed wavelength of the blend in our +25\,d spectrum yields $z$ = 0.041\,$\pm$\,0.001. This is consistent with the redshift of 0.042 reported by \citet{2013CBET3616} for the host galaxy of SN 2013eq. As a further test we performed a series of cross correlations using SNID \citep[Supernova Identification,][]{Blondin2007}. Suggested matches were scrutinized and the selected results span a range of redshifts consistent with our result. We therefore adopt $z$ = 0.041\,$\pm$\,0.001 as redshift of the SN.

\subsection{Photometry}
\label{section:Photometry}

\begin{figure}
  \centering
    \includegraphics[width=\columnwidth]{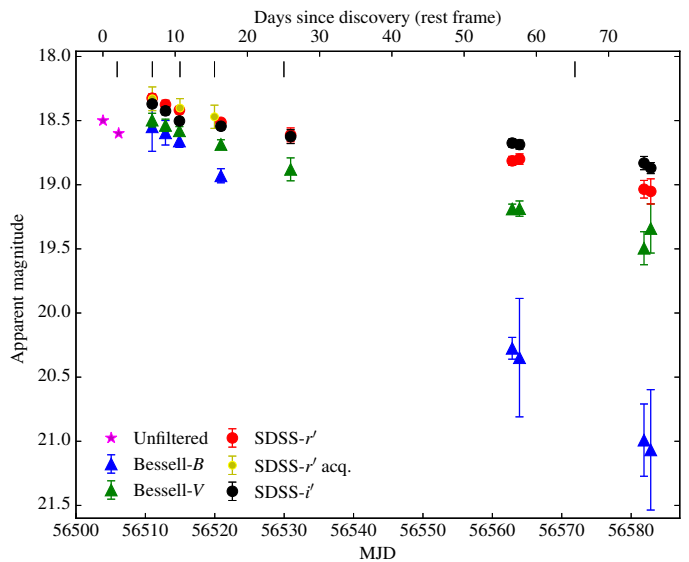}
    \caption{Bessell-$B$, $V$, SDSS-$r'$, $i'$ light curves of SN 2013eq. The vertical ticks on the top mark the epochs of the observed spectra. The unfiltered magnitudes are from \citep{2013CBET3616}.}
    \label{figure:SN2013eq_lightcurve}
\end{figure}

SN 2013eq was likely observed shortly after peaking in the optical bands (the light curve clearly shows a decline before settling onto the plateau), and the initial magnitudes of 18.548\,$\pm$\,0.030 and 18.325\,$\pm$\,0.022 measured in Bessell-$B$ and SDSS-$r'$, respectively, are presumably very close to the maximum in these bands. 
The light curves initially decline at relatively steep rates of 
3.5\,$\pm$\,0.1 mag/100\,d in Bessell-$B$,
2.1\,$\pm$\,0.2 mag/100\,d in Bessell-$V$,
2.1\,$\pm$\,0.2 mag/100\,d in SDSS-$r'$, and
3.4\,$\pm$\,0.1 mag/100\,d in SDSS-$i'$ 
until about 10 to 15 days after discovery. Then the decline rates slow down to
0.95\,$\pm$\,0.02 mag/100\,d in Bessell-$V$,
0.17\,$\pm$\,0.02 mag/100\,d in SDSS-$r'$, and
0.60\,$\pm$\,0.03 mag/100\,d in SDSS-$i'$ between $\sim$25\,d and $\sim$55\,d after discovery, whilst the Bessel-$B$ light curve displays no break.
Type II-P SNe display a plateau phase with almost constant brightness after a short (or sometimes negligible) initial decline \citep[e.g.][]{Anderson2014a}. Typically the plateau phase lasts up to 100\,d before the light curve drops onto the radioactive tail. This transition was, however, not observed for SN 2013eq.
Table \ref{table:photometry} in the appendix shows the log of imaging observations as well as the final calibrated magnitudes. The light curve is presented in Figure \ref{figure:SN2013eq_lightcurve}.

\subsection{Spectroscopy}
\label{section:Spectroscopy}

\begin{figure}
  \centering
    \includegraphics[width=\columnwidth]{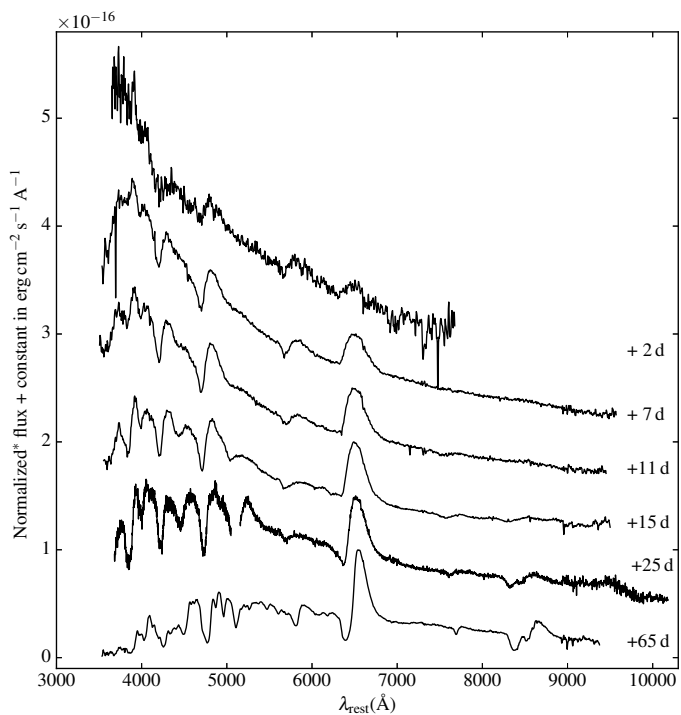}
    \caption{SN 2013eq spectroscopy. The first (+2\,d) spectrum is the classification spectrum \citep{2013CBET3616,Tomasella2014}. $^*$Flux normalized to the maximum H$\alpha$ flux for better visibility of the features. The exact normalizations are: 
flux/1.9 for the +1\,d and +7\,d spectrum;
flux/2.0 for +11\,d;
flux/2.2 for +15\,d;
flux/2.1 for +25\,d;
flux/2.3 for +65\,d.
}
    \label{figure:SN2013eq_spectra}
\end{figure}

Table \ref{table:journal_spectra} in the appendix shows the journal of spectroscopic observations. In addition to the spectra obtained for our study we also included the publicly available classification spectrum\footnote{Classification spectra from the Asiago Transient Classification Program \citep{Tomasella2014} are publicly available at http://graspa.oapd.inaf.it/cgi-bin/output\_class.cgi?sn=2011.} obtained on 2013 July 31 and August 1 \citep{2013CBET3616}. The fully reduced and calibrated spectra of SN 2013eq are presented in Figure \ref{figure:SN2013eq_spectra}. They are corrected for reddening ($E(B - V)_{\mathrm{tot}}$ = 0.096\,mag) and redshift ($z$ = 0.041). 

The strongest feature in the spectra is H$\alpha$, which would be matched by the corresponding features at about 4850\,\AA, 4300\,\AA\ and 4100\,\AA\ in the blue to be H$\beta$, H$\gamma$ and H$\delta$. 
In the classification spectrum, H$\alpha$ and H$\beta$ profiles are visible though relatively weakly. They become stronger at later epochs and we can discern also H$\gamma$, and H$\delta$. 
We can also discern a feature at about 5900\,\AA\ which is typically assigned to He\,{\sc i} $\lambda$5876. \citet{Leonard2002a} claim that it evolves into a blend with Na\,{\sc i}\,D at later epochs in SN 1999em.

Finally, we can observe weak lines of iron between 4000 and 5500\,\AA. In particular Fe\,{\sc ii}\,$\lambda$5169 which is visible in the spectra from +11\,d on. Fe\,{\sc ii}\,$\lambda$5018 is visible only in the +65\,d spectrum. At this epoch we also see weak lines around 4450\,\AA\ and 4860\,\AA\ which have been attributed to a blend of Fe\,{\sc ii}, Ba\,{\sc ii} and Ti\,{\sc ii} in SN 1999em by \citet{Leonard2002a}.
The pseudo-equivalent width (pEW) of the Fe\,{\sc ii}\,$\lambda$5018 feature can be used as
a proxy for the progenitor metallicity \citep{Dessart2014}. For SN~2013eq, we measure a pEW$_{\mathrm{Fe\,II}\,\lambda 5018}$ = $-$13.4\,$\pm$\,0.4\,\AA\, in our 65\,d spectrum, that would place it in the 0.1-0.4\,$Z_{\astrosun}$ range, provided the 15\,$M_\odot$ models are an  appropriate choice for SN~2013eq. Interestingly, in their study of the potentially very low metallicity ($Z \lesssim 0.1\,Z_\odot$) Type II-P supernova, LSQ13fn, \citet{Polshaw2016} find that the SCM relation is violated.

\section{Methods}
\label{section:Methods}

\subsection{The expanding photosphere method}
\label{section:The_expanding_photosphere_method}

As mentioned earlier, the EPM was originally suggested by \citet{Kirshner1974} and in the past decades 
a number of improvements and variations have been presented. In this section, however, we will only outline the most basic principle of the EPM, laying the foundation for the more detailed derivations in Section \ref{section:EPM_at_high_redshifts} regarding the application of the EPM at higher redshifts.

The photospheric angular size $\vartheta$ of a Type II supernova of redshifts $z \ll 1$ can be described as\footnote{Note that most publications that make use of the EPM leave out the factor of ``2'' in Eqn. \ref{equation:EPM_vartheta}. Although it eventually cancels out in the final distance result (see Equation \ref{equation:EPM_D_2v_vartheta_tt0}) in the interest of completeness and correctness, we will preserve this factor in Eqn. \ref{equation:EPM_vartheta}. \label{footnote:Factor2}}:
\begin{equation}
\label{equation:EPM_vartheta}
\vartheta = 2\frac{R}{D} = 2\sqrt{\frac{f_\lambda}{\zeta_\lambda^2\pi B_\lambda(T)}} ,
\end{equation}
where $R$ is the photospheric radius, $D$ the distance to the SN, $f_\lambda$ the observed flux density, $\zeta_\lambda$ the distance correction factor or dilution factor, 
and $B_\lambda(T)$ the Planck function evaluated at observed photospheric temperature $T$. 
$\zeta_\lambda$ is derived from model atmospheres and is used to correct dilution effects of scattering atmospheres meaning that a SN will not emit as a perfect black body.  
Approximately one day after explosion the SN achieves a state of homologous expansion and the photospheric radius of the SN at a certain time $t$ is 
\begin{equation}
\label{equation:EPM_R}
R = v(t - t_0) + R_0 .
\end{equation}
$v$ is the photospheric expansion velocity of the SN, 
$t_0$ is the time of explosion and $R_0$ the initial radius. Compared to the extent of the ejecta $R_0$ becomes negligible very soon. It should be noted that neglecting $R_0$ can introduce a small error when applying Equation \ref{equation:EPM_R} at epochs within the first 1-2 weeks after explosion, where -- depending on the photospheric velocities and the initial radius of a particular SN -- $R_0$ might still be in the 10\% range of the photospheric radius $R$. Combining equations \ref{equation:EPM_vartheta} and \ref{equation:EPM_R} the relation between the photospheric angular size $\vartheta$, the photospheric expansion velocity $v$, measured at time $t$ can be expressed as
\begin{equation}
\label{equation:EPM_D_2v_vartheta_tt0}
D = \frac{2v}{\vartheta}(t-t_0)
\end{equation}
This means that with a minimum of two measurements spaced in time of $\vartheta$ and $v$, Equation \ref{equation:EPM_D_2v_vartheta_tt0} can be solved for the distance $D$ and time of explosion $t_0$.

\subsection{EPM at high redshifts}
\label{section:EPM_at_high_redshifts}

The comparison of the flux with the angular size of an object forms the cornerstone on which the concept of the EPM rests (cf. Equation \ref{equation:EPM_vartheta}). However, it is advisable to be prudent when dealing with SNe at non-negligible redshifts, where the ``angular distance'' and the ``luminosity distance'' differ by a factor of $(1+z)^2$ \citep[e.g.][]{Rich2001}: 
\begin{equation}
\label{Dtheta_DL}
D_\theta = \frac{D_L}{(1+z)^2} . 
\end{equation}
While at $z = 0.01$ a factor of $(1+z)^2$ results only in a $\sim$2\% discrepancy, this effect increases quadratically and at a redshift of only 0.05 the difference between luminosity distance and angular distance is already $\sim$10\%. 

Bearing in mind our derived redshift of $z$ = 0.041\,$\pm$\,0.001 for SN~2013eq, relativistic effects cannot be ignored. We therefore re-derive the basic equation for the EPM for non-negligible redshifts. In the following ``$\star$'' will be used to denote the SN rest frame, while ``$\Diamond$'' will be used for variables in the observed frame.

The luminosity distance to the SN, $D_L$, can be expressed as 
\begin{equation}
\label{equation:f_L_DL_monochromatic}
f^\mathrm{dered}_{\Diamond\lambda_\Diamond} = \frac{L_{\lambda_\star}}{4\pi D_L^2} \frac{1}{(1+z)} ,
\end{equation}
where $f^\mathrm{dered}_{\Diamond \lambda_\Diamond}$ is the observed energy flux in the observed wavelength interval $\Delta\lambda_\Diamond$, corrected for galactic and extragalactic extinction and $L_{\lambda_\star}$ is the total monochromatic luminosity in the rest frame wavelength interval $\Delta\lambda_\star$. The factor $(1+z)$ takes into account that  $f^\mathrm{dered}_{\Diamond\lambda_\Diamond}$ and $L_{\lambda_\star}$ are given in different coordinate systems.
$L_{\lambda_\star}$ can also be expressed in terms of the monochromatic radiation emitted in all directions: $L_{\lambda_\star} = \zeta_{\lambda_\star}^2 \pi B_{\lambda_\star}(T_\star) \cdot 4\pi R_\star^2$, where $B_{\lambda_\star}(T_\star)$ is the Planck function evaluated at the photospheric temperature $T_\star$ of the SN and $4\pi R_\star^2$ is the surface area of the SN with radius $R_\star$. $\zeta_{\lambda_\star}$ is the dilution factor in the SN rest frame. 
Equation \ref{equation:f_L_DL_monochromatic} can therefore be rewritten as: 
\begin{equation}
\label{equation:RD_sqrt_fderedz1_zeta2piBT}
\frac{R_\star}{D_L} = \sqrt{\frac{f^\mathrm{dered}_{\Diamond\lambda_\Diamond} (1 + z)}{\zeta_{\lambda_\star}^2 \pi B_{\lambda_\star} (T_\star)}} .
\end{equation}

The photospheric angular size, or angular separation of the photosphere, $\theta$\footnote{Note that we use the symbol ``$\theta$'' to denote the actual angular size, while the symbol ``$\vartheta$'' denotes the {\em approximation} of the angular size that is only valid at low redshifts (as in Equations \ref{equation:EPM_vartheta} and \ref{equation:EPM_D_2v_vartheta_tt0}).} of a SN can be expressed as 
\begin{equation}
\label{equation:theta_RDtheta}
\theta = 2\frac{R_\star}{D_{\theta}} ,
\end{equation}
where $D_\theta$ is the ``angular distance''.

When dealing with non-negligible redshifts, the two terms $\frac{R_\star}{D}$ in Equations \ref{equation:RD_sqrt_fderedz1_zeta2piBT} and \ref{equation:theta_RDtheta}, are obviously not the same. While Equation \ref{equation:EPM_vartheta} is indeed valid for $z \ll 1$ Equations \ref{Dtheta_DL}, \ref{equation:RD_sqrt_fderedz1_zeta2piBT} and \ref{equation:theta_RDtheta} can be combined to derive the correct relation between $\theta$ and the observed flux:
\begin{equation}
\theta = 2\,(1+z)^2 \sqrt{\frac{f^\mathrm{dered}_{\Diamond\lambda_\Diamond} (1+z)}{\zeta_{\lambda_\star}^2 \pi B_{\lambda_\star} (T_\star)}} 
= (1+z)^2 \theta^\dag ,
\label{equation:theta_highz}
\end{equation}
where we defined $\theta^\dag := \theta/(1+z)^2$. Even though $\theta^\dag$ is not an ``angular size'' in the mathematical sense, it corresponds\footnote{aside from the aforementioned factor of ``2''(see Footnote \ref{footnote:Factor2})} to the ``angular size'', $\vartheta$, that was utilized for EPM in previous publications \citep[e.g.][]{Schmidt1994a,Jones2009}.
Equation \ref{equation:EPM_D_2v_vartheta_tt0} thus transforms to
\begin{equation}
\label{equation:DL_2v_theta_tt0}
D_\theta = \frac{2v}{\theta}(t^\star-t_0^\star) = \frac{2v}{(1+z)^2 \theta^\dag}(t^\star-t^\star_0) ,
\end{equation}
or
\begin{equation}
\label{equation:DL_2v_thetadag_tt0}
D_L = \frac{2v}{\theta^\dag}(t^\star-t^\star_0) ,
\end{equation}
where $t^\star$ is the time in the SN rest frame.

After careful examination of the various equations finding their way into the final distance result, let us draw attention to a few particular points. 
\begin{itemize}
\item First, the term $f^\mathrm{dered}_{\Diamond\lambda_\Diamond} (1+z)$ in Equation \ref{equation:theta_highz} can be transformed to $f^\mathrm{dered}_{\star\lambda_\star}$ in the SN rest frame by applying a $K$-correction to observed flux. This was already recognized by \citet[][see their Equation 6]{Schmidt1994a} who also note that their $\theta$ (which corresponds to $^1/_2\,\theta^\dag$ in this paper) is not an ``angular  size'' as in \citet{Wagoner1977}. Consequently, their distance result has to be interpreted as a luminosity distance. 

\item Second, in order to determine the correct angular size, $\theta$, of objects at non-negligible redshifts the factor of $(1+z)^2$ has to be taken into account. 

\item Third, when calculating the luminosity distance instead of the angular distance the factors of $(1+z)^2$ in Equations \ref{Dtheta_DL} and \ref{equation:theta_highz} ``cancel'' each other out, resulting in Equation \ref{equation:DL_2v_thetadag_tt0} which is basically identical to the formulation that is commonly used (see Equation \ref{equation:EPM_D_2v_vartheta_tt0}). In short, the use of $\theta$ will result in an angular distance, while the use of $\theta^\dag$ will result in a luminosity distance.  
This is a vital distinction that has, to our best knowledge, hitherto been ignored. Accordingly, EPM distance results in literature that follow a formulation similar to Equation \ref{equation:EPM_D_2v_vartheta_tt0} -- the correct high-redshift formulation of which is Equation \ref{equation:DL_2v_thetadag_tt0} -- should be regarded as luminosity distances. 
\end{itemize}

We want to emphasize that in order to correctly apply the EPM also to SNe at non-negligible redshifts the only correction to be made (compared to the low redshift EPM) is the $K$-correction of the observed flux. This implies the use of $\theta^\dag$ in Equation \ref{equation:DL_2v_thetadag_tt0} and also means that previously published applications of the EPM are correct (regarding this matter) if the resulting distances are seen as luminosity distances and $K$-corrections were either applied or negligible.

\subsection{The standardized candle method}
\label{section:The_standardized_candle_method}

The SCM for Type II SNe was first suggested by \citet{Hamuy2002a}. Here however, we follow the approach of \citet{Nugent2006}, who were the first to apply the SCM to SNe at redshifts of up to z\,$\sim$\,0.3. The basic concept of the SCM is briefly outlined in the following.

Equation 1 in \citet{Nugent2006} describes a correlation between the rest frame $I$-band magnitude, $M_I$, the rest frame $(V - I)$-colour and the expansion velocity at 50 days after explosion:
\begin{equation}
\label{equation:SCM_Nugent2006}
M_{I_{50}} = -\alpha\, \mathrm{log}_{10}\left(\frac{ v_{50,\mathrm{Fe\,\textsc {ii} }} }{5000}\right) - 1.36 \left[ (V - I)_{50} - (V - I)_0 \right] + M_{I_0} ,
\end{equation}
with $\alpha = 5.81$, $M_{I_0} = -17.52$ (for an $H_0$ of 70\,km\,s$^{-1}$\,Mpc$^{-1}$) and $(V - I)_0 = 0.53$. An advantage of this formulation is that no reddening correction is required for the observed photometry which can introduce an additional error if the host galaxy reddening is unknown. This works reasonably well under the assumption that the extinction laws are similar in most galaxies. Therefore the observed magnitudes only need to be transformed to the rest frame which can be easily attended to by applying a $K$-correction.
Consequently, equation \ref{equation:SCM_Nugent2006} can be adopted for local as well as more distant SNe. A practical limitation may be the spectral coverage of the rest frame $I$ band at higher redshift needed for a precise $K$-correction.

The expansion velocity is typically estimated using the Fe\,{\sc ii}\,$\lambda$5169 line \citep[e.g.][]{Hamuy2002a,Nugent2006,Poznanski2009}. As spectroscopic data 50 days after explosion might not always be available, \citet{Nugent2006} explored the time dependence of the Fe\,{\sc ii}\,$\lambda$5169 velocity and found that at 50\,d after explosion it can be estimated using the following relation \citep[][equation 2]{Nugent2006}:
\begin{equation}
\label{equation:v50_Nugent2006}
v_{50} = v(t^\star) \left( \frac{t^\star}{50} \right)^{0.464\,\pm\,0.017} ,
\end{equation}
where $v(t^\star)$ is the Fe\,{\sc ii}\,$\lambda$5169 velocity at time $t^\star$ after explosion.

\section{Results and discussion}
\label{section:Results_and_discussion}

Measuring the distance to SN 2013eq using the expanding photosphere or standardized candle method, requires a number of preparatory steps, like deriving the colour temperatures and velocities (for EPM) or the magnitudes and velocities at 50 days after explosion (for SCM). The EPM additionally relies on the use of appropriate dilution factors\footnote{Strictly, the term ``correction factors'' would be more appropriate as the calculations incorporate departures from a blackbody spectrum above and beyond a non-zero albedo.}. \citet{Eastman1996} calculated model atmospheres for Type II SNe and derived a set of dilution factors for the filter combinations \{$BV$\}, \{$VI_C$\}, \{$BVI_C$\} and \{$JHK$\}. \citet{Hamuy2001} later re-calculated the dilution factors for a different photometric system and expanded the number of filter combinations to \{$BV$\}, \{$VI$\}, \{$BVI$\}, \{$VZ$\}, \{$VJ$\}, \{$VH$\}, \{$VK$\}, and \{$JHKN$\} using the same atmospheric models. \citet{Dessart2005} also computed dilution factors for the filter sets \{$BV$\}, \{$VI$\}, \{$BVI$\} and \{$JHK$\}, based on a large set of photospheric-phase models of Type II SNe. The $R$-band is typically excluded due to the strong H$\alpha$ contribution.

In the case of SN 1999em, the EPM distance derived by \citet{Leonard2002a} using the dilution factors from \citet{Hamuy2001} was $\sim$\,30\,\% shorter than the Cepheid distance to the host galaxy of SN 1999em, NGC 1637 \citep[11.7\,$\pm$\,1.0\,Mpc,][]{Leonard2003}. This discrepancy was resolved by \citet{Dessart2006} who applied the dilution factors reported in \citet{Dessart2005}; these are systematically larger than those of \citet{Hamuy2001}. This resulted in a distance estimate to  $\sim$1.7\,\% of the Cepheid value.

Here, we will use the dilution factors given by both \citet{Hamuy2001} and \citet{Dessart2005}, for the filter combination $BVI$, respectively, so as to include all available data for SN 2013eq in our EPM distance estimate.

\subsection{Temperature evolution}
\label{Section:Temperature_evolution}

Ideally, an estimate of the {\em photospheric} temperature should be used for the EPM. In practice however, it is difficult to directly measure the photospheric temperature. Consequently, the {\em colour} temperature is commonly used for the EPM as an estimator for the photospheric temperature \citep[e.g.][]{Hamuy2001,Leonard2002a,Dessart2005,Jones2009}. While these are conceptionally different, this should be a reasonable approximation. In particular, the dilution factors presented in either \citet{Hamuy2001} or \citet{Dessart2005} are functions of specific colour temperatures. In our case we use the $B$, 	$V$, and $I$ photometry to estimate the colour temperature as to be consistent with the $BVI$ dilution factors.

This in turn requires us to transform the SDSS-$i$ band photometry to the Johnson-Cousins Filter System used in \citet{Hamuy2001} and \citet{Dessart2005}. Taking into consideration that the EPM requires the flux in the SN rest frame we first calculate $K$-corrections by using the SN 2013eq spectroscopy and the {\sc snake} code (SuperNova Algorithm for $K$-correction Evaluation) within the S3 package \citep{Inserra2016}. For the SDSS-$i$ filter we determine the $K$-correction to the rest frame Johnson-Cousins $I$ band. 
Then, the uncorrected observed photometry is interpolated to the epochs of spectroscopic observations, and subsequently dereddened and $K$-corrected. Eventually, the adjusted $B$, $V$, and $I$-band magnitudes are converted into physical fluxes. The colour temperature at each epoch is derived by fitting a black body curve to the deduced fluxes and effective wavelengths of the corresponding filters.

We carried out ancillary blackbody fits and added or subtracted the uncertainties in all possible combinations. The standard deviation of the resulting range of temperatures was taken to be a conservative estimate of the uncertainty in the temperature. We adopted this approach in order to make full use of all available information across all bands. 
 
The results are presented in Table \ref{table:Temperature_evolution}.

\begin{table*}
  \caption{Interpolated rest frame photometry and temperature evolution of SN 2013eq}
  \label{table:Temperature_evolution}
  \centering
  \begin{tabular}{c c     c            c                   c                    c                     c                     }
  \hline 
\multirow{2}{*}{Date} & \multirow{2}{*}{MJD} & Epoch$^*$  & $B^{**}$           & $V^{**}$          & $I^{**}$            & $T_{\star,BVI}$      \\
             &           & rest frame (d) & mag                &   mag              &  mag                &    K                \\
\hline 
  2013-08-01 & 56505.93  &  +1.97     & 17.95\,$\pm$\,0.27 & 18.09\,$\pm$\,0.09 & 17.71\,$\pm$\,0.07 & $>$12000             \\
  2013-08-06 & 56511.00  &  +6.84     & 18.15\,$\pm$\,0.19 & 18.25\,$\pm$\,0.10 & 17.81\,$\pm$\,0.06 & 12498\,$\pm$\,7532   \\
  2013-08-10 & 56514.99  & +10.67     & 18.29\,$\pm$\,0.15 & 18.36\,$\pm$\,0.10 & 17.96\,$\pm$\,0.06 & 11888\,$\pm$\,4175   \\
  2013-08-15 & 56519.96  & +15.44     & 18.43\,$\pm$\,0.12 & 18.45\,$\pm$\,0.11 & 17.99\,$\pm$\,0.07 & 10622\,$\pm$\,2347   \\
  2013-08-25 & 56529.96  & +25.05     & 18.73\,$\pm$\,0.12 & 18.55\,$\pm$\,0.11 & 18.15\,$\pm$\,0.08 &  8580\,$\pm$\,1050   \\
  2013-10-06 & 56571.90  & +65.34     & 19.91\,$\pm$\,0.21 & 18.93\,$\pm$\,0.12 & 18.24\,$\pm$\,0.07 &  5318\,$\pm$\, 242   \\
  \hline  
  \end{tabular}
  \\[1.5ex]
  \flushleft
  $^*$Rest frame epochs (assuming a redshift of 0.041) with respect to the first detection on 56503.882 (MJD). $^{**}$Magnitudes in the Johnson-Cousins Filter System, $K$-corrected and amended for dust extinction. For the first epoch the data are insufficient to estimate a temperature. We therefore derived a lower limit of 12000\,K using a black body fit to the interpolated photometry.
\end{table*}

\subsection{Velocities}

The knowledge of the line velocities in SN 2013eq is crucial to both the EPM and the SCM. The velocities are measured using {\sc iraf} by fitting a Gaussian function to the minima of the various lines. The uncertainty in the velocity determination is presumed to be in the 5\,\% range.  

\citet{Dessart2005} show that the velocity measured from the Fe\,{\sc ii}  $\lambda$5169 absorption matches the photospheric velocity within 5-10\%. However, they also point out that this line is only visible at later epochs. 
\citet{Leonard2002a} and \citet{Leonard2002b} argue that lines such as Fe\,{\sc ii} $\lambda\lambda$5018,5169 will overestimate the photospheric velocities and that instead weak Fe\,{\sc ii} lines such as  $\lambda\lambda$4629,4670,5276,5318 give more accurate results. An overestimation of the photospheric velocities would lead to distances that are too large for both the EPM and the SCM. 

Whilst a precise stipulation of the photospheric velocity is indeed desirable we have to bear in mind that in the case of SN 2013eq, many of the weaker lines favored by \citet{Leonard2002a} and \citet{Leonard2002b} for the EPM are only visible in the +65\,d spectrum. However, the EPM requires measurements of the photospheric velocity for no less than two epochs. This leaves only three possible Fe\,{\sc ii} lines that are visible in more than one spectrum of SN 2013eq: Fe\,{\sc ii} $\lambda$5018, Fe\,{\sc ii} $\lambda$5169 and the blended line Ba\,{\sc ii} / Fe\,{\sc ii} $\lambda$6147. For the purposes of the EPM an average of the measured absorption velocities from these three lines will be used (see Table \ref{table:EPM_Quantities}). 

For the SCM we infer the velocity at 50 days after explosion by means of the Fe\,{\sc ii}\,$\lambda$5169 velocities and Equation \ref{equation:v50_Nugent2006} (see Section \ref{section:The_SCM_distance_to_SN_2013eq} for more details).

\subsection{The EPM distance to SN 2013eq}

Having performed all necessary measurements we are now ready to wade into the final steps towards calculating an EPM distance for SN 2013eq. As mentioned above, we use the $BVI$ dilution factors as given by \citet[][which are based on the dilution factors calculated by \citealt{Eastman1996}]{Hamuy2001} and \citet{Dessart2005}. Both groups find that the dilution factor essentially is a function of the colour temperature that can be described as $\zeta(T_S) = \sum_i a_{S,i} (10^4\,\mathrm{K}/T_S)^i$ with S representing the filter subset (in our case \{$BVI$\}). The parameters $a_{S,i}$ are given in Appendix C, Table 14 for \citet{Hamuy2001} and Table 1 for \citet{Dessart2005}. 

In order to apply Equation \ref{equation:theta_highz} to SN 2013eq we rewrite it as 
\begin{equation}
\theta^\dag = 2 \sqrt{\frac{f^\mathrm{dered}_{\star,F} }{\zeta_{\star,BVI}^2(T_{\star,BVI}) \pi B(\lambda_{\star,\mathrm{eff}_F},T_{\star,BVI})}} ,
\label{equation:theta_highz_SN2013eq}
\end{equation}
where $\lambda_{\star,\mathrm{eff}_F}$ is the effective wavelength of the corresponding Filter, $F$, in the rest frame. Note that we use $\theta^\dag$ instead of $\theta$ in our calculations in order not to introduce an additional error on the final result due to the uncertainty of the redshift.

\begin{figure*}[t!]
   \centering
   \begin{subfigure}[t]{0.49\textwidth}
      \includegraphics[width=\columnwidth]{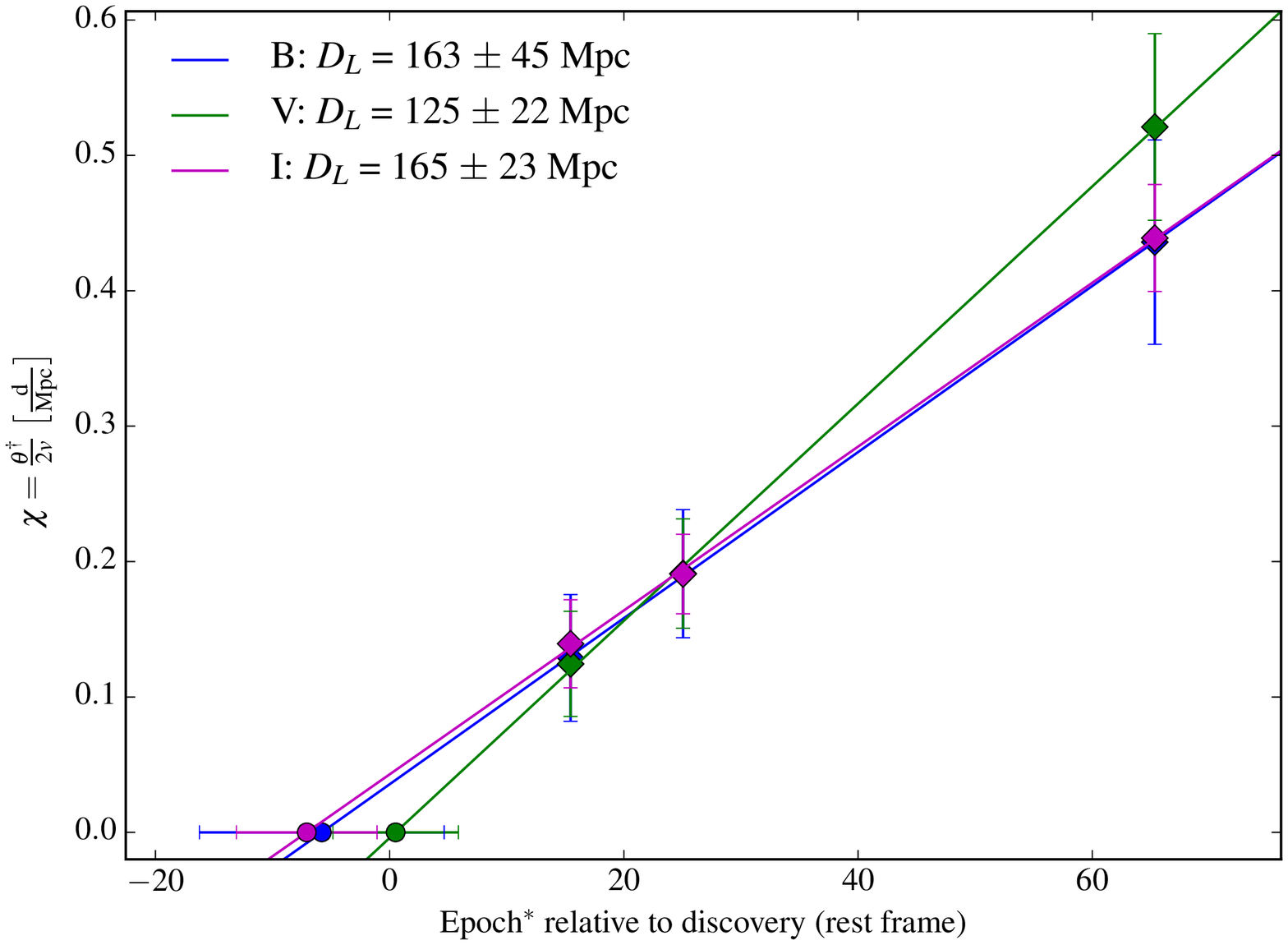}
      \label{figure:Distances_vaverage_BVI_H01}
   \end{subfigure}%
   \begin{subfigure}[t]{0.49\textwidth}
      \includegraphics[width=\columnwidth]{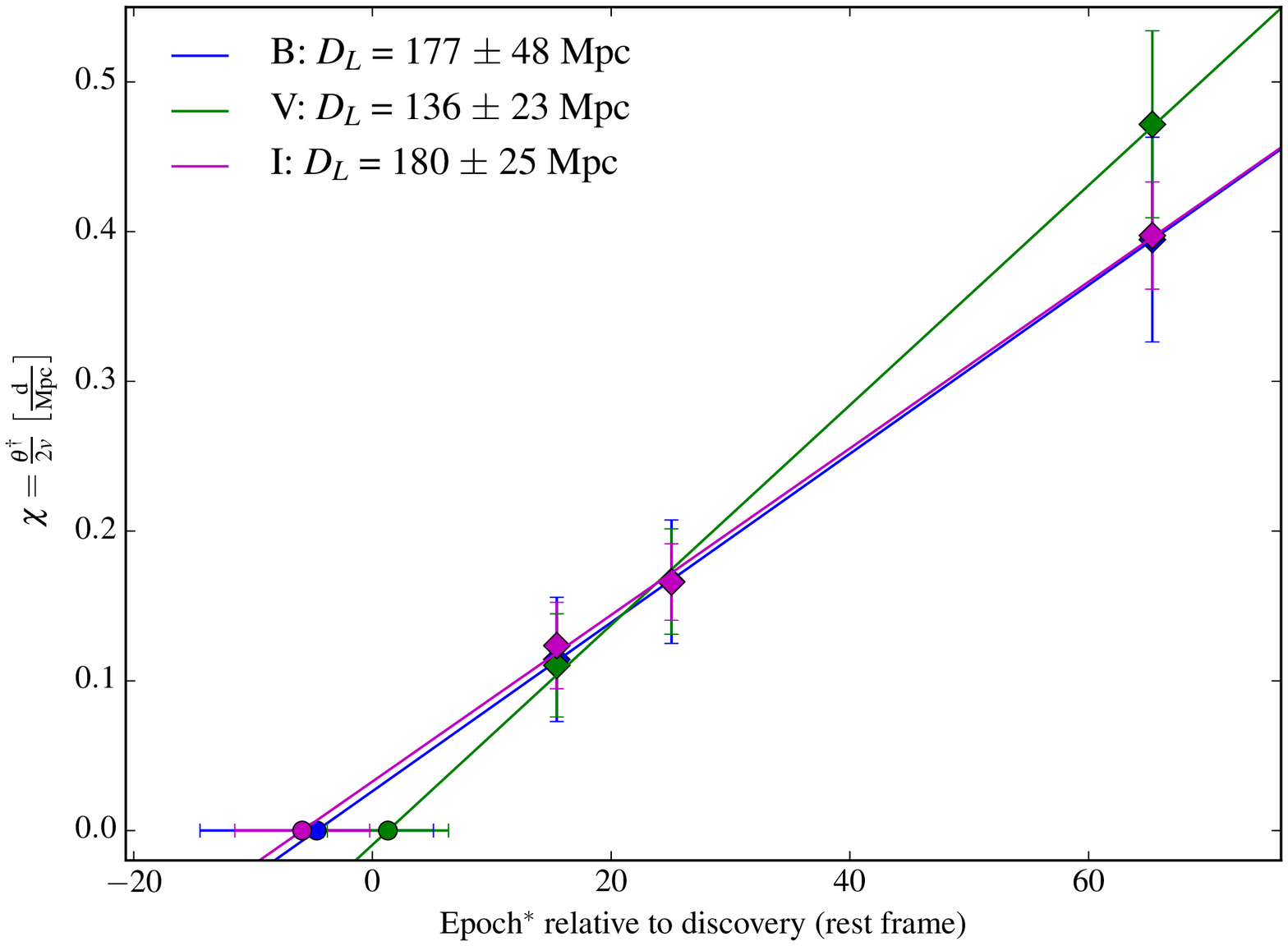}
      \label{figure:Distances_vaverage_BVI_D05}
   \end{subfigure}%
   \caption{Distance fit for SN 2013eq using $\zeta_{BVI}$ as given in \citet{Hamuy2001} (left panel) and \citet{Dessart2005} (right panel). The diamond markers denote values of $\chi$ through which the fit is made; circle markers depict the resulting epoch of explosion.
}
   \label{figure:Distances_vaverage}
\end{figure*}

Finally, rearranging Equation \ref{equation:DL_2v_thetadag_tt0} gives us:
\begin{equation}
\label{equation:EPMFIT_chi_tt0_D}
\chi := \frac{\theta^\dag}{2v} = \frac{t^\star-t^\star_0}{D_L} .
\end{equation}
For each filter $B$, $V$, and $I$, values of $\chi$ at the corresponding epochs are fitted linearly using $D_L$ and $t^\star_0$ as parameters. 
The errors are estimated by executing complementary fits through the same values of $\chi$, but adding or subtracting the uncertainties in all possible combinations. The standard deviations of the resulting range in distances as well as epochs of explosion are then employed as conservative estimates of their respective uncertainties. 
The results are presented in Figure \ref{figure:Distances_vaverage} and Table \ref{table:EPM_distance_t0} for each set of dilution factors from \citet{Hamuy2001} and \citet{Dessart2005}, respectively. 

Note that while the derived distances and explosion epochs in the $B$- and $I$-band are very similar, the $V$-band distance is about 24\% smaller than the $I$-band distance for both sets of dilution factors. Similar effects have also been observed by other groups performing the EPM. \citet{Hamuy2001}, for example, find a difference of up to 20\% in their distances when applying the EPM to SN 1999em for varying filter combinations and attribute these disparities to systematic errors in the dilution corrections. \citet{Jones2009} apply the EPM to 12 SNe and find varying results for for different filter combinations for all their SNe. In the case of SN 1999em the distances reported by \citet{Jones2009} differ by almost 26\% between the ${BV}$ and the ${VI}$ filter combinations. 
It stands to reason that the discrepancies appearing when applying the EPM for different filter combinations are reflected also when applying it to single filters instead of filter combinations. 

This disagreement could be due to the use of tabulated dilution factors. \citet{Dessart2006} find that when using models tailored to SN~1999em, the EPM distances derived in the different band passes are in accord with each other. An advantage of using tailored models is that the dilution factors can be evaluated alongside other EPM parameters (such as the colour temperature), and can thereby be adjusted for differences between individual SNe. For instance, our observations (\S \ref{section:Spectroscopy}) imply a subsolar metallicity for SN~2013eq, but the dilution factors are only available for solar metallicity models.
Nevertheless, such tailored approaches remain currently unviable even for small sample sizes, and will be even more so for the significantly larger samples that will become available in the near future.

Our final errors on the distance are in the range of $\sim$\,12\% and take into account the uncertainties from the magnitudes, the redshift, the $K$-corrections, the velocities and the host galaxy reddening. These are propagated through all calculations and build the basis for the uncertainties of the flux, the colour temperature, and the angular size. 
In an attempt to estimate the global errors of their distances \citet{Jones2009} perform 1000 Monte Carlo simulations varying over parameters such as photometry, redshift, foreground \& host-galaxy extinction, line expansion velocity, photospheric velocity conversion as well as dilution factors and find errors between 10 and 41\% for distances to the SNe in their sample.
\citet{Leonard2002b} follow a similar approach in calculating a set of simulations while varying over the velocities, magnitudes and dilution factors. They find statistical distance errors for SN 1999em of only a few percent, however they point out that the error of the EPM distance to SN 1999em is likely dominated by systematic errors as the results derived using dilution factors for different filter combinations vary by as much as 19\%. 
\citet{Dessart2006} discuss potential error sources in detail. In particular, they note that the uncertainty in the angular size, propagated from the error in the flux, is temperature dependent, and that an uncertainty in the flux has a larger influence on the final error than an uncertainty in $E(B-V)$. This stems from the fact that extinction effects on the temperature estimate and the measure of the angular size will compensate each other to some extent. This has also been discussed previously by \citet{Schmidt1992}, \citet{Eastman1996}, and \citet{Leonard2002a}.

\begin{table*}
  \caption{EPM Quantities for SN 2013eq}
  \label{table:EPM_Quantities}
  \centering
  \begin{tabular}{c           c                           c                         c                                   c                    c                    c                   c           c }
  \hline 
  \multirow{2}{*}{Date}       & \multirow{2}{*}{MJD}      & Epoch$^*$               & Averaged v & \multirow{2}{*}{$\theta_B^\dag\,\times\,10^{12}$} & \multirow{2}{*}{$\theta_V^\dag\,\times\,10^{12}$}    & \multirow{2}{*}{$\theta_I^\dag\,\times\,10^{12}$} & \multicolumn{2}{c}{Dilution factor} \\
                              &                           & rest frame              & km\,s$^{-1}$                          &             &                    &                   & $\zeta_{BVI}$ & reference \\
\hline 
\multirow{2}{*}{2013-08-15} & \multirow{2}{*}{56519.96} & \multirow{2}{*}{+15.44} & \multirow{2}{*}{6835\,$\pm$\,244} & 4.9\,$\pm$\,1.8 &  4.8\,$\pm$\,1.5 & 5.3\,$\pm$\,1.2 &  0.41 & H01  \\
                            &                           &                         &                                   & 4.4\,$\pm$\,1.6 &  4.2\,$\pm$\,1.3 & 4.7\,$\pm$\,1.1 &  0.53 & D05   \\  
\multirow{2}{*}{2013-08-25} & \multirow{2}{*}{56529.96} & \multirow{2}{*}{+25.05} & \multirow{2}{*}{5722\,$\pm$\,202} & 6.1\,$\pm$\,1.5 &  6.1\,$\pm$\,1.3 & 6.1\,$\pm$\,0.9 &  0.43 & H01  \\
                            &                           &                         &                                   & 5.3\,$\pm$\,1.3 &  5.3\,$\pm$\,1.1 & 5.3\,$\pm$\,0.8 &  0.59 & D05   \\
\multirow{2}{*}{2013-10-06} & \multirow{2}{*}{56571.90} & \multirow{2}{*}{+65.34} & \multirow{2}{*}{3600\,$\pm$\,104} & 8.8\,$\pm$\,1.5 & 10.5\,$\pm$\,1.4 & 8.8\,$\pm$\,0.8 &  0.75 & H01  \\
                            &                           &                         &                                   & 8.0\,$\pm$\,1.4 &  9.5\,$\pm$\,1.2 & 8.0\,$\pm$\,0.7 &  0.92 & D05   \\
  \hline  
  \end{tabular}
  \\[1.5ex]
  \flushleft
  $^*$Rest frame epochs (assuming a redshift of 0.041) with respect to the first detection on 56503.882 (MJD). H01: \citet{Hamuy2001}; D05: \citet{Dessart2005}. See also Figure \ref{figure:Distances_vaverage}. 
\end{table*}

\begin{table*}
  \caption{EPM distance and explosion time for SN 2013eq}
  \label{table:EPM_distance_t0}
  \centering
  \begin{tabular}{c   c                         c                 c                   c                c                  c                  }
  \hline         
  Dilution            & \multirow{2}{*}{Filter} &  $D_L$         & Averaged $D_L$ & $t^\star_0$           & Average $t^\star_0$     & $t^\Diamond_0$      \\
  Factor              &                         &  Mpc            &      Mpc      &    days$^{*}$   &    days$^{*}$    & mjd                  \\
\hline        
 \multirow{3}{*}{H01} &  $B$                    & 163\,$\pm$\,45 &                &  \phs5.8\,$\pm$\,10.5 &                  &                      \\
                      &  $V$                    & 125\,$\pm$\,22 & 151\,$\pm$\,18 & $-$0.5\,$\pm$\, 5.4 & 4.1\,$\pm$\,4.4  & 56499.6\,$\pm$\,4.6  \\
                      &  $I$                    & 165\,$\pm$\,23 &                &  \phs7.1\,$\pm$\, 6.0 &                  &                       \\ 
 \hline                
 \multirow{3}{*}{D05} &  $B$                    & 177\,$\pm$\,48 &                & \phs4.7\,$\pm$\,9.8 &                  &                       \\
                      &  $V$                    & 136\,$\pm$\,23 & 164\,$\pm$\,20 & $-$1.3\,$\pm$\,5.1 & 3.1\,$\pm$\,4.1  & 56500.7\,$\pm$\,4.3   \\
                      &  $I$                    & 180\,$\pm$\,25 &                & \phs5.9\,$\pm$\,5.6 &                  &                        \\ 
  \hline  
  \end{tabular}
  \\[1.5ex]
  \flushleft
  $^*$Rest frame days before discovery on 56503.882 (MJD). H01: \citet{Hamuy2001}; D05: \citet{Dessart2005}. See also Figure \ref{figure:Distances_vaverage}.  
\end{table*}

\subsection{The SCM distance to SN 2013eq}
\label{section:The_SCM_distance_to_SN_2013eq}

Compared to the EPM the utilization of the SCM is somewhat less laborious. As mentioned in Section \ref{section:The_standardized_candle_method}, to begin with, the magnitudes used for the SCM have to be transformed to the SN rest frame. We therefore calculate $K$-corrections by using the SN 2013eq spectroscopy and the {\sc snake} code (SuperNova Algorithm for $K$-correction Evaluation) within the S3 package \citep{Inserra2016}. For the SDSS-$i'$ band we select the $K$-corrections to the rest frame Johnson-Cousins $I$ Filter. These values are then interpolated to an epoch of 50\,d post-explosion. 

We next need to derive the magnitude $I_{50}$ and colour $(V - I)_{50}$ i.e., 50\,d  after explosion. We therefore fit low-order polynomials to the uncorrected photometry and subsequently transform it to the rest frame by applying the $K$-corrections. The expansion velocity is determined using the relation in equation \ref{equation:v50_Nugent2006}. 
We then use our measured Fe\,{\sc ii}\,$\lambda$5169 velocities to estimate an appropriate value for 50\,d. 

We repeated the above procedure three times using the epoch of explosion derived via EPM with the dilution factors from \citet{Hamuy2001} and \citet{Dessart2005}. We additionally, estimated the explosion epoch by utilizing the average rise time for SNe II-P of 7.0\,$\pm$\,0.3 as given by \citet{Gall2015} assuming that SN~2013eq was discovered close to maximum. This is likely a valid assumption given its spectral and photometric evolution. Applying, Equation \ref{equation:SCM_Nugent2006} -- which essentially describes the relation between the $I$-band magnitude, the $(V - I)$-colour and the expansion velocity 50 days after explosion -- for each of the three cases, gives us $M_{I_{50}}$. Finally, the distance modulus can be calculated, $\mu$ = $I_{50} - M_{I_{50}}$, which in turn can be converted into a luminosity distance via $D_L$ [Mpc] = $10^{\mu/5-5}$\,Mpc.

Note that the uncertainty in the explosion epoch will introduce an error in all quantities derived at 50 days after explosion. An earlier explosion epoch will ultimately result in a larger distance. This has been discussed also by \citet{Nugent2006}, who found that the explosion date uncertainty has the largest impact on the final error compared to other contributions, and also by \citet{Poznanski2009} who, in contrast, find that their results vary only little with the explosion epoch arguing that the magnitudes and colours are relatively constant during the plateau phase. 
We find that even though the explosion epoch for SN 2013eq by applying the EPM has relatively large uncertainties, this contributes only little to the uncertainties of the $V$- and $I$-band magnitudes and consequently also the ($V - I$) colour at 50 days after explosion. These are of the same order as the original uncertainties in the photometry. The time of explosion uncertainty is, however, more significant when determining the expansion velocity at 50 days, although the error in the Fe\,{\sc ii}\,$\lambda$5169 velocities and the intrinsic error in Equation \ref{equation:v50_Nugent2006} also contribute to the total error.

The uncertainty in the redshift plays an almost negligible role. For completeness we did however propagate its error when accounting for time dilation. Note that \citet{Hamuy2002a} find peculiar motions in nearby galaxies ($cz < 3000$\,km\,s$^{-1}$) contribute significantly to the overall scatter in their Hubble diagram; however this is not a relevant issue for SN 2013eq. 

The final uncertainties in the distance modulus and the distance are propagated from the errors in $M_{I_{50}}$, $v_{50,\mathrm{Fe\,\textsc {ii} }}$ and $(V - I)_{50}$.
The derived distance moduli and luminosity distances as well as the intermediate results are given in Table \ref{table:SCM_distance}.

\begin{table*}
  \caption{SCM quantities and distance to SN 2013eq}
  \label{table:SCM_distance}
  \centering
  \begin{tabular}{c c            c                     c                    c                 c                     c                  }
  \hline
  Estimate & $t^\Diamond_0$               &  $V^{*}_{50}$       &  $I^{*}_{50}$     & $v_{50}$        & $\mu$              & $D_L$              \\
  of $t_0$ via   & mjd                 &   mag               &   mag              &    km\,s$^{-1}$ & mag                &    Mpc             \\
\hline        
 EPM -- H01       & 56499.6\,$\pm$\,4.6 & 19.05\,$\pm$\,0.09 & 18.39\,$\pm$\,0.04 & 4880\,$\pm$\,760 & 36.03\,$\pm$\,0.43 & 160\,$\pm$\,32 \\
 EPM -- D05       & 56500.7\,$\pm$\,4.3 & 19.06\,$\pm$\,0.09 & 18.39\,$\pm$\,0.04 & 4774\,$\pm$\,741 & 35.98\,$\pm$\,0.42 & 157\,$\pm$\,31 \\
 Rise time -- G15 & 56496.6\,$\pm$\,0.3 & 19.03\,$\pm$\,0.05 & 18.39\,$\pm$\,0.04 & 5150\,$\pm$\,353 & 36.13\,$\pm$\,0.20 & 168\,$\pm$\,16 \\

  \hline  
  \end{tabular}
  \\[1.5ex]
  \flushleft
  $^{*}K$-corrected magnitudes in the Johnson-Cousins Filter System. H01: \citet{Hamuy2001}; D05: \citet{Dessart2005}. See also Figure \ref{figure:Distances_vaverage}. 
\end{table*}

\subsection{Comparison of EPM and SCM distances}

An inspection of Table \ref{table:EPM_distance_t0} reveals that the two EPM luminosity distances derived using the dilution factors from \citet{Hamuy2001} and \citet{Dessart2005} give consistent values. This is no surprise, bearing in mind that the dilution factors from \citet{Hamuy2002a} and \citet{Dessart2005} applied for SN 2013eq differ by only 18-27\,\% (see Table \ref{table:EPM_Quantities}). Similarly, the resulting explosion epochs are also consistent with each other. 

Likewise, the SCM distances calculated utilizing the times of explosion found via EPM and the dilution factors from either \citet{Hamuy2002a} or \citet{Dessart2005} (see Table \ref{table:SCM_distance}), as well as by adopting the average SN II-P rise time as given by \citet{Gall2015}, are consistent not only with each other but also with the EPM results. 

It is remarkable how close our outcomes are within the errors to the distance of 176\,Mpc calculated from the redshift of SN 2013eq with the simple formula $D = cz/H_0$ (for $H_0$ = 70\,km\,s$^{-1}$\,Mpc$^{-1}$). While this is of course no coincidence for the SCM-distances (which are based on $H_0$ = 70\,km\,s$^{-1}$\,Mpc$^{-1}$), the EPM-distance is completely independent as to any assumptions concerning the Hubble constant. This is particularly encouraging, considering the scarcity of data points for our fits stemming mostly from the difficulty of measuring the velocities of weak iron lines in our spectra. It seems that both the SCM and the EPM are surprisingly robust techniques to determine distances even at non-negligible redshifts where high cadence observations are not always viable.

\section{Conclusions}
\label{section:Conclusions}

We presented optical light curves and spectra of the Type II-P SN 2013eq. It has a redshift of $z$ = 0.041\,$\pm$\,0.001 which inspired us to embark on an analysis of relativistic effects when applying the expanding photosphere method to SNe at non-negligible redshifts. 

We find that for the correct use of the EPM to SNe at non-negligible redshifts, the observed flux needs to be converted into the SN rest frame, e.g. by applying a $K$-correction. In addition, the angular size, $\theta$, has to be corrected by a factor of $(1+z)^2$ and the resulting EPM distance will be an angular distance. 
However, when using a modified version of the angular size $\theta^\dag = \theta/(1+z)^2$ the EPM can be applied in the same way as has previously been done for small redshifts, with the only modification being a $K$-correction of the observed flux. The fundamental difference is that this will result in a luminosity distance instead of an angular distance.

For the SCM we follow the approach of \citet{Nugent2006}, who outline its use for SNe at cosmologically significant redshifts. Similar to the EPM their formulation of the high redshift SCM requires the observed magnitudes to be transformed into the SN rest frame, which in practice corresponds to a $K$-correction.

We find EPM luminosity distances of $D_L$ = 151\,$\pm$\,18\,Mpc and $D_L$ = 164\,$\pm$\,20\,Mpc as well as times of explosions of 4.1\,$\pm$\,4.4\,d and 3.1\,$\pm$\,4.1\,d before discovery (rest frame), by using the dilution factors in \citet{Hamuy2001} and \citet{Dessart2005}, respectively. Assuming that SN 2013eq was discovered close to maximum light this would result in rise times that are in line with those of local SNe II-P \citep{Gall2015}. With the times of explosions derived via the EPM -- having used the dilution factors from either \citet{Hamuy2001} or \citet{Dessart2005} -- we find SCM luminosity distances of $D_L$ = 160\,$\pm$\,32\,Mpc and $D_L$ = 157\,$\pm$\,31\,Mpc. By utilizing the average rise time of SNe II-P as presented in \citet{Gall2015} to estimate the epoch of explosion we find an independent SCM distance of $D_L$ = 168\,$\pm$\,16\,Mpc.

The luminosity distances derived using different dilution factors as well as either EPM or SCM are consistent with each other. Considering the scarcity of viable velocity measurements it is encouraging that our results lie relatively close to the expected distance of $\sim$\,176\,Mpc calculated from the redshift of SN 2013eq. Conversely, the EPM distances can be used to calculate the Hubble constant, which (using $D = cz/H_0$) results in $H_0$ = 83\,$\pm$\,10\,km\,s$^{-1}$\,Mpc$^{-1}$ and $H_0$ = 76\,$\pm$\,9\,km\,s$^{-1}$\,Mpc$^{-1}$ applying the dilution factors from \citet{Hamuy2001} and \citet{Dessart2005}, respectively. These are consistent with the latest results from \citet[][$H_0$ = 73.0\,$\pm$\,1.8\,km\,s$^{-1}$\,Mpc$^{-1}$]{Riess2016}.

With current and upcoming transient surveys, it appears to be only a matter of time until statistically significant numbers of SNe II-P become available also at non-negligible redshifts. Consequently, the promise of yielding sound results will turn the EPM and SCM into increasingly important cosmological tools, provided that the requisite follow-up capabilities are in place.

\section*{Acknowledgements} 

We are grateful to an anonymous referee for their comments and suggestions.
EEEG, BL, ST, and WH acknowledge support for this work by the Deutsche Forschungsgemeinschaft through the TransRegio project TRR33 ‘The Dark Universe’.
RK acknowledges support from STFC via ST/L000709/1. 
Based in part on observations made with the Gran Telescopio Canarias (GTC2007-12ESO, PI:RK), installed in the Spanish Observatorio del Roque de los Muchachos of the Instituto de Astrofísica de Canarias, on the island of La Palma. 
The William Herschel Telescope and its service programme are operated on the island of La Palma by the Isaac Newton Group in the Spanish Observatorio del Roque de los Muchachos of the Instituto de Astrofísica de Canarias.
The Liverpool Telescope is operated on the island of La Palma by Liverpool John Moores University in the Spanish Observatorio del Roque de los Muchachos of the Instituto de Astrofisica de Canarias with financial support from the UK Science and Technology Facilities Council. 
This research has made use of the NASA/IPAC Extragalactic Database (NED) which is operated by the Jet Propulsion Laboratory, California Institute of Technology, under contract with the National Aeronautics and Space Administration.

\bibliography{SN2013eq_paper} \bibliographystyle{aa}

\appendix

\section{Additional tables}

Table \ref{table:sequence_stars} lists the local sequence stars that were used to calibrate the photometry of SN 2013eq (see also Section \ref{section:Data_reduction}).

\begin{table*}
 \caption{Optical sequence stars}
 \label{table:sequence_stars}
 \centering
 \begin{tabular}{c c  c		         c            c                    c            c                    c            c                    c            c             }
 \hline                 
 \multirow{2}{*}{Star} & \multirow{2}{*}{R.A.} & \multirow{2}{*}{Dec.} & \multicolumn{2}{c}{Bessell $B$} & \multicolumn{2}{c}{Bessell $V$} & \multicolumn{2}{c}{SDSS $r'$} & \multicolumn{2}{c}{SDSS $i'$} \\ 
      &              &              &  mag       & error              &  mag       & error              &  mag       & error              &  mag       & error        \\
\hline                 
 1    & 17:33:16.920 & +36:28:13.48 &  19.251    &  0.044             &  18.677    &  0.033             &  18.293    &  0.030             &  18.079    &  0.021       \\
 2    & 17:33:19.218 & +36:28:06.82 &  19.294    &  0.044             &  18.098    &  0.026             &  17.259    &  0.019             &  16.603    &  0.016       \\
 3    & 17:33:17.518 & +36:27:49.13 &  20.254    &  0.083             &  18.973    &  0.041             &  18.208    &  0.030             &  17.584    &  0.014       \\
 4    & 17:33:15.701 & +36:27:28.09 &  19.875    &  0.065             &  18.546    &  0.030             &  17.719    &  0.022             &  16.695    &  0.010       \\
 5    & 17:33:10.460 & +36:27:17.35 &  19.307    &  0.043             &  18.755    &  0.036             &  18.385    &  0.033             &  18.186    &  0.023       \\
 6    & 17:33:02.387 & +36:26:44.13 &  17.538    &  0.017             &  17.062    &  0.013             &  16.745    &  0.014             &  16.581    &  0.010       \\
 7    & 17:33:08.137 & +36:26:01.99 &  19.939    &  0.073             &  18.697    &  0.037             &  17.917    &  0.025             &  17.078    &  0.011       \\
 8    & 17:33:15.256 & +36:26:13.71 &  19.780    &  0.056             &  18.483    &  0.032             &  17.660    &  0.022             &  16.280    &  0.009       \\
 9    & 17:33:15.912 & +36:26:24.68 &  21.074    &  0.182             &  19.675    &  0.073             &  19.043    &  0.054             &  18.520    &  0.034       \\
 10   & 17:33:15.047 & +36:26:38.23 &  18.303    &  0.022             &  17.821    &  0.017             &  17.521    &  0.018             &  17.369    &  0.014       \\
 11   & 17:33:23.409 & +36:26:08.63 &  20.283    &  0.088             &  19.232    &  0.057             &  18.429    &  0.029             &  17.852    &  0.019       \\
 12   & 17:33:28.011 & +36:29:49.18 &  18.462    &  0.023             &  17.901    &  0.021             &  17.516    &  0.017             &  17.353    &  0.039       \\
 13   & 17:33:22.442 & +36:30:27.64 &  17.512    &  0.016             &  17.203    &  0.013             &  16.913    &  0.034             &  16.895    &  0.011       \\
 14   & 17:33:22.348 & +36:29:45.53 &  19.704    &  0.057             &  18.468    &  0.036             &  17.670    &  0.021             &  17.179    &  0.013       \\
 15   & 17:33:17.825 & +36:29:49.07 &  20.870    &  0.163             &  19.656    &  0.077             &  18.870    &  0.049             &  17.988    &  0.023       \\
 16   & 17:33:17.123 & +36:29:05.82 &  19.063    &  0.038             &  18.337    &  0.025             &  17.898    &  0.021             &  17.664    &  0.015       \\
 17   & 17:33:13.687 & +36:29:11.64 &  18.720    &  0.027             &  17.954    &  0.021             &  17.422    &  0.017             &  17.154    &  0.013       \\
 18   & 17:33:10.398 & +36:29:47.99 &  18.899    &  0.033             &  18.147    &  0.022             &  17.660    &  0.020             &  17.403    &  0.013       \\
 19   & 17:33:03.204 & +36:29:20.77 &  17.583    &  0.015             &  17.170    &  0.014             &  16.899    &  0.013             &  16.812    &  0.011       \\
 20   & 17:32:58.976 & +36:28:54.90 &  18.400    &  0.028             &  17.713    &  0.023             &  17.253    &  0.015             &  17.075    &  0.014       \\
 \hline  
 \end{tabular}
 \\[1.5ex]
 \flushleft
\end{table*}

Tables \ref{table:photometry} and \ref{table:journal_spectra} show the journals of imaging and spectroscopic observations, respectively. In addition to the spectra obtained for our study we also included the classification spectrum obtained on 2013 July 31 and August 1 with the Asiago Faint Object Spectrograph and Camera (AFOSC) mounted on the Asiago 1.82-m Copernico Telescope \citep{2013CBET3616,Tomasella2014}.

\begin{table*}
  \caption{Photometric observations}
  \label{table:photometry}
  \centering
  \begin{tabular}{l l    c            c        c        c               c        c        c               c           }
  \hline               
\multirow{2}{*}{Date} & \multirow{2}{*}{MJD} & Rest frame & \multicolumn{3}{c}{Bessell $B$} & \multicolumn{3}{c}{Bessell $V$} & \multirow{2}{*}{Telescope} \\ 
             &          & epoch$^*$  & mag    & error  & $K$-corr$^{**}$ & mag    & error  & $K$-corr$^{**}$ &            \\
\hline                        
  2013-08-06 & 56510.98 &   +6.82    & 18.548 & 0.191  & -0.017\,$\pm$\,0.007 & 18.496 & 0.053  & -0.064\,$\pm$\,0.007 & LT         \\ 
  2013-08-08 & 56512.90 &   +8.66    & 18.595 & 0.095  & -0.004\,$\pm$\,0.007 & 18.537 & 0.048  & -0.069\,$\pm$\,0.007 & LT         \\ 
  2013-08-10 & 56514.90 &  +10.58    & 18.659 & 0.050  &  0.009\,$\pm$\,0.007 & 18.579 & 0.032  & -0.072\,$\pm$\,0.007 & LT         \\ 
  2013-08-16 & 56520.89 &  +16.34    & 18.930 & 0.055  &  0.046\,$\pm$\,0.006 & 18.686 & 0.037  & -0.062\,$\pm$\,0.007 & LT         \\ 
  2013-08-26 & 56530.93 &  +25.98    &   -    &  -     &   -                  & 18.880 & 0.089  &  0.001\,$\pm$\,0.007 & LT         \\ 
  2013-09-27 & 56562.86 &  +56.66    & 20.275 & 0.085  &  0.253\,$\pm$\,0.006 & 19.189 & 0.038  &  0.042\,$\pm$\,0.007 & LT         \\ 
  2013-09-28 & 56563.91 &  +57.66    & 20.348 & 0.462  &  0.257\,$\pm$\,0.006 & 19.186 & 0.060  &  0.043\,$\pm$\,0.007 & LT         \\ 
  2013-10-16 & 56581.85 &  +74.90    & 20.991 & 0.282  &  0.316\,$\pm$\,0.009 & 19.495 & 0.129  &  0.066\,$\pm$\,0.007 & LT         \\ 
  2013-10-17 & 56582.82 &  +75.83    & 21.067 & 0.469  &  0.319\,$\pm$\,0.009 & 19.341 & 0.192  &  0.067\,$\pm$\,0.007 & LT         \\ 
  \hline               
  \hline               
\multirow{2}{*}{Date} & \multirow{2}{*}{MJD} & Rest frame & \multicolumn{3}{c}{SDSS $r'$}    & \multicolumn{3}{c}{SDSS $i'$}    & \multirow{2}{*}{Telescope} \\ 
             &          & epoch$^*$  & mag    & error & $K$-corr$^{**}$  & mag    & error & $K$-corr$^{**}$  &            \\
\hline                          
  2013-08-06 & 56510.98 &   +6.82    & 18.325 & 0.033 & 0.027\,$\pm$\,0.007 & 18.370 & 0.024 & -0.080\,$\pm$\,0.007 & LT         \\ 
  2013-08-07 & 56511.00 &   +6.84    & 18.330 & 0.093 & 0.027\,$\pm$\,0.007 &  -     &  -    &  -                   & GTC        \\ 
  2013-08-08 & 56512.90 &   +8.66    & 18.375 & 0.034 & 0.045\,$\pm$\,0.007 & 18.424 & 0.031 & -0.091\,$\pm$\,0.007 & LT         \\ 
  2013-08-10 & 56514.90 &  +10.58    & 18.418 & 0.029 & 0.060\,$\pm$\,0.007 & 18.503 & 0.027 & -0.102\,$\pm$\,0.007 & LT         \\ 
  2013-08-10 & 56514.98 &  +10.66    & 18.402 & 0.073 & 0.061\,$\pm$\,0.007 &  -     &  -    &  -                   & GTC        \\ 
  2013-08-15 & 56519.96 &  +15.44    & 18.470 & 0.090 & 0.086\,$\pm$\,0.007 &  -     &  -    &  -                   & GTC        \\ 
  2013-08-16 & 56520.89 &  +16.34    & 18.513 & 0.028 & 0.088\,$\pm$\,0.007 & 18.544 & 0.028 & -0.128\,$\pm$\,0.007 & LT         \\ 
  2013-08-26 & 56530.93 &  +25.98    & 18.611 & 0.054 & 0.080\,$\pm$\,0.007 & 18.625 & 0.054 & -0.155\,$\pm$\,0.007 & LT         \\ 
  2013-09-27 & 56562.86 &  +56.66    & 18.814 & 0.036 & 0.159\,$\pm$\,0.007 & 18.675 & 0.032 & -0.162\,$\pm$\,0.007 & LT         \\ 
  2013-09-28 & 56563.91 &  +57.66    & 18.800 & 0.041 & 0.161\,$\pm$\,0.007 & 18.687 & 0.035 & -0.161\,$\pm$\,0.007 & LT         \\ 
  2013-10-16 & 56581.85 &  +74.90    & 19.035 & 0.069 & 0.196\,$\pm$\,0.007 & 18.831 & 0.052 & -0.108\,$\pm$\,0.007 & LT         \\ 
  2013-10-17 & 56582.82 &  +75.83    & 19.052 & 0.098 & 0.198\,$\pm$\,0.007 & 18.870 & 0.043 & -0.104\,$\pm$\,0.007 & LT         \\ 
  \hline  
  \end{tabular}
  \\[1.5ex]
  \flushleft
  The tabulated magnitudes are given ``as observed'', i.e. neither corrected for dust extinction or $K$-corrected. $^*$Rest frame epochs (assuming a redshift of 0.041) with respect to the first detection on 56503.882 (MJD). 
  $^{**}K$-corrections were calculated by first using the SN 2013eq spectroscopy and the {\sc snake} code (SuperNova Algorithm for $K$-correction Evaluation) within the S3 package \citep{Inserra2016}. These values were then interpolated to all epochs of photometric observations. The $K$-corrections are not included in the tabulated photometry. For the SDSS-$r'$ and SDSS-$i'$ band we calculate the $K$-correction to the rest frame Johnson-Cousins $R$ and $I$ band. 
LT = Liverpool Telescope; GTC = Gran Telescopio CANARIAS. 
\end{table*}

\begin{table*}
  \caption{Journal of spectroscopic observations}
  \label{table:journal_spectra}
  \centering
  \begin{tabular}{c c     c            c              c            c                               }
  \hline 
\multirow{2}{*}{Date} & \multirow{2}{*}{MJD} & Epoch$^*$  & Wavelength   & Resolution & \multirow{2}{*}{Telescope+Instrument} \\
             &           & rest frame & range in \AA &  \AA       &                               \\
\hline 
  2013-08-01 & 56505.93  &  +1.97     & 3654 $-$ \phn7692  &  13.0      & Asiago 1.82-m Telescope+AFOSC+GR04 \\
  2013-08-06 & 56511.00  &  +6.84     & 3538 $-$ \phn9590  &  17.4      & GTC+OSIRIS+R300B              \\
  2013-08-10 & 56514.99  & +10.67     & 3510 $-$ \phn9471  &  17.3      & GTC+OSIRIS+R300B              \\
  2013-08-15 & 56519.96  & +15.44     & 3558 $-$ \phn9519  &  16.6      & GTC+OSIRIS+R300B              \\
  2013-08-25 & 56529.96  & +25.05     & 3400 $-$ 10600 & 3.8/6.9    & WHT+ISIS+R158R/R300B          \\
  2013-10-06 & 56571.90  & +65.34     & 3538 $-$ \phn9394  &  16.9      & GTC+OSIRIS+R300B              \\
  \hline  
  \end{tabular}
  \\[1.5ex]
  \flushleft
  $^*$Rest frame epochs (assuming a redshift of 0.041) with respect to the first detection on 56503.882 (MJD). The first (+2\,d) spectrum is the classification spectrum \citep{2013CBET3616,Tomasella2014}. The resolution of the GTC and WHT spectra was determined from the FWHM of the O\,{\sc i} $\lambda$5577.34 sky line.
\end{table*}

\end{document}